\documentclass{moriond}

\def\be{\begin{equation}}
\def\ee{\end{equation}}
\def\bea{\begin{eqnarray}}
\def\eea{\end{eqnarray}}

\newcommand{\Photo}{\includegraphics[height=35mm]{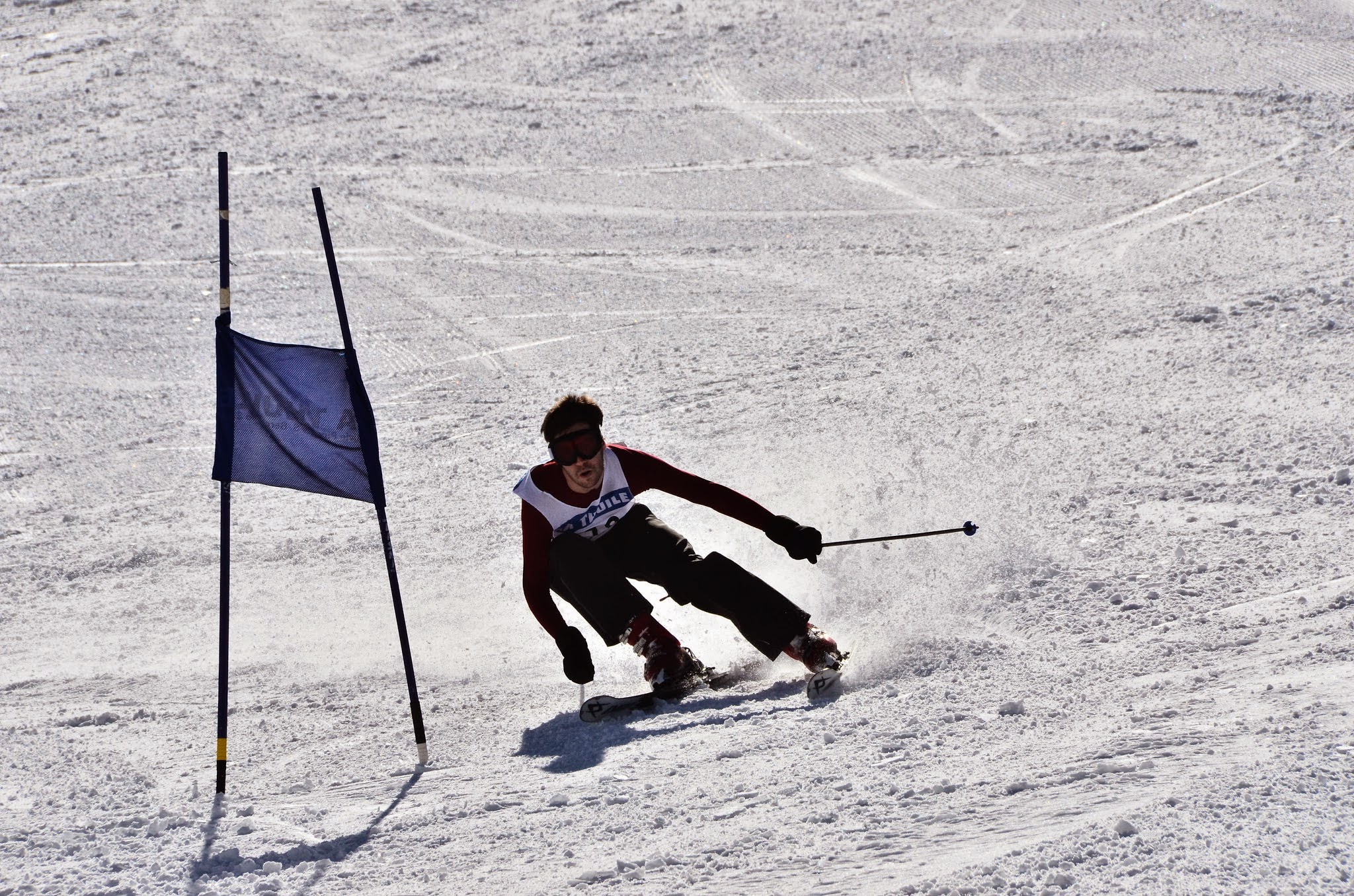}}

\usepackage{xspace}
\providecommand{\HERAFitter}{{\texttt{HERA\-Fitter}}\xspace}

\usepackage{graphicx}
\usepackage{booktabs}
\usepackage{subfigure}
\usepackage{placeins}
\usepackage{amsmath}

\begin{document}
\vspace*{4cm}
\title{HERAFitter - An Open Source framework to determine PDFs}

\author{Stefano Camarda}
\address{DESY - Notkestra{\ss}e 85
D-22607 Hamburg Germany\\
On behalf of the HERAFitter Collaboration.}

\maketitle\abstracts{
  The \HERAFitter{} project provides a framework for the determination
  of parton distribution functions (PDFs), and tools for assessing the impact of new
  data on PDFs.
  In this contribution, \HERAFitter{} is used for a QCD analysis of the legacy measurements
  of the $W$-boson charge asymmetry and of the
  $Z$-boson production cross sections, performed at the Tevatron
  collider in Run II by the D0 and CDF collaborations.
  The Tevatron measurements are included in a PDF fit performed at
  next-to-leading order, and compared to the predictions obtained using
  other PDF sets from different groups. The measurements are in good agreement with NLO QCD
  theoretical predictions. The Tevatron data provide significant constraints on the $d$-valence quark distribution.
}

\section{Introduction}

\begin{figure}
  \begin{center}
    \includegraphics[width=0.6\textwidth]{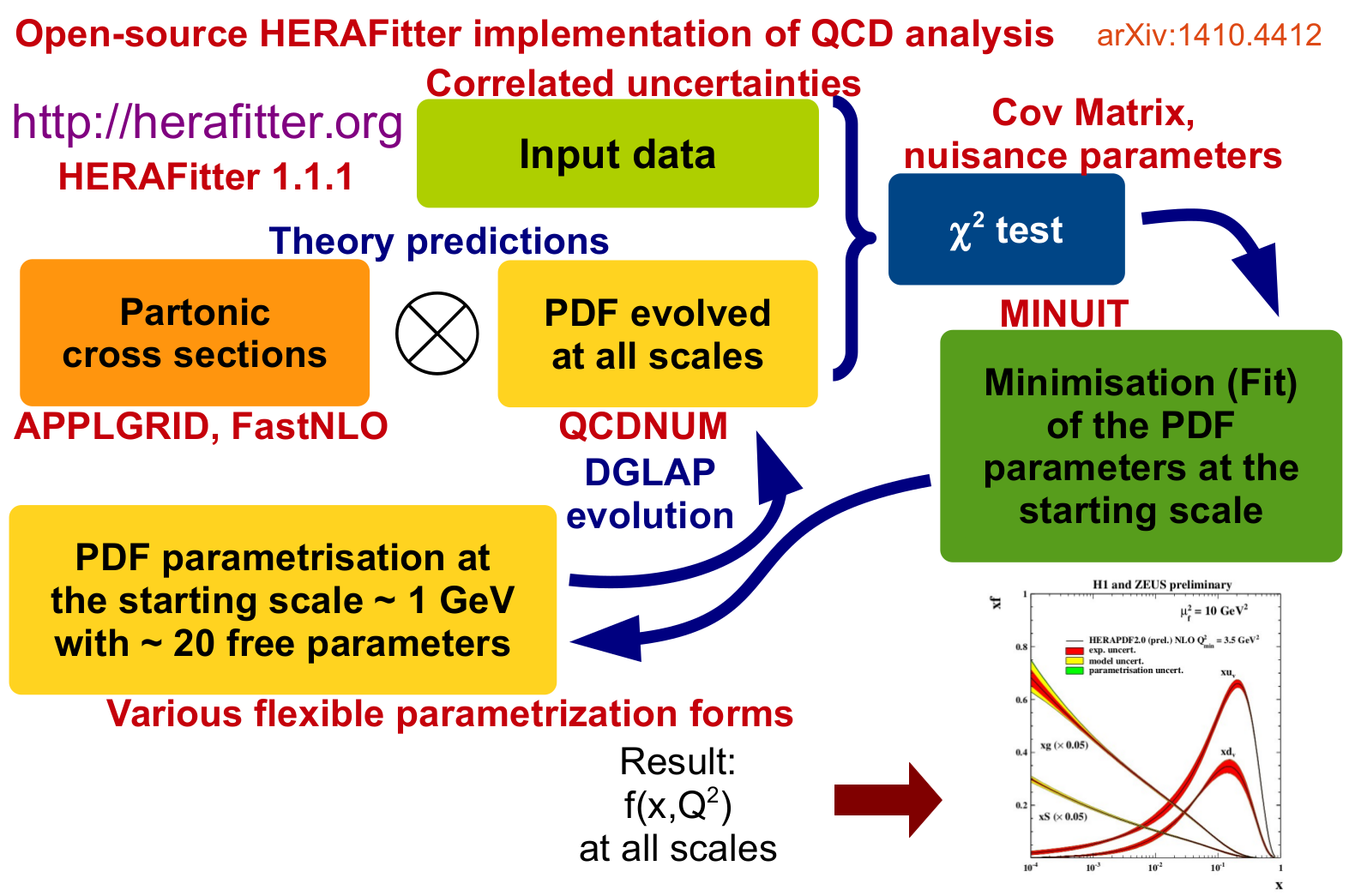}
  \end{center}
    \caption{Schematic representation of a PDF fit in \HERAFitter{}.}
	\label{fig:PDFfit}
\end{figure}

According to the factorisation theorem, cross sections in hadron
collisions are calculated by convoluting short distance partonic
reactions with parton distribution functions (PDFs).
Discovery of physics beyond the Standard Model at hadron colliders relies on the precise
knowledge of the proton structure. Moreover, PDFs are among the dominant
uncertainties for the measurement of the $W$ mass, and for $gg \to H$
production.
\HERAFitter~\cite{Alekhin:2014irh} provides a framework for the investigation of various
methodologies in PDF fits, and tools for assessing the impact of new
data on PDFs. It is widely used by the LHC experiments to improve the
sensitivity of new measurements to PDFs. Full information about the
project, downloads and documentation can be found at
\emph{herafitter.org}.
A schematic view of a PDF fit, as implemented in \HERAFitter{}, is shown in Fig.~\ref{fig:PDFfit}

In this contribution, \HERAFitter{} is used to perform a QCD analysis
of the Tevatron Run II legacy measurements of the $W$-boson charge asymmetry and of
the $Z$-boson production cross sections~\cite{Camarda:2015zba}.
At the Tevatron proton-antiproton collider, the production of $W$ and $Z$
bosons is dominated by valence-quark interactions.
Whereas the primary source of information on the proton PDFs comes
from deep-inelastic scattering (DIS), Drell-Yan production of $W$ and
$Z$ bosons in proton-antiproton collisions can provide additional
information, particularly on the $d$-valence quark PDFs.

\section{Tevatron $W$ and $Z$ measurements and QCD settings}
The most recent measurements of $W$-boson charge asymmetry and
$Z$-boson inclusive production performed in Run II of the Tevatron
collider are considered. They include the $Z$-boson
differential cross section as a function of rapidity, measured by
D0~\cite{Abazov:2007jy}; the $Z$-boson differential
cross section as a function of rapidity, measured by CDF
~\cite{Aaltonen:2010zza}; the charge asymmetry of
muons as a function of rapidity in $W \to \mu \nu$ decays, measured by
 D0~\cite{Abazov:2013rja}; the $W$-boson charge
asymmetry as a function of rapidity, measured by CDF~\cite{Aaltonen:2009ta}; the $W$-boson charge asymmetry as a
function of rapidity, measured by
 D0~\cite{Abazov:2013dsa}.
Besides the Tevatron $W$- and $Z$-boson measurements,
the HERA I combined measurements of the inclusive 
DIS neutral- and charged-current cross sections measured by the H1 and ZEUS
experiments~\cite{Aaron:2009aa} are used.

In general, the correlation model of the experimental uncertainties
recommended by the Tevatron experiments is adapted and followed in the QCD
analysis, with the exception of the experimental systematic
uncertainties related to trigger and lepton identification
efficiencies, which are treated as uncorrelated bin-to-bin.

The QCD analysis and PDF extraction is performed with the open-source
framework \HERAFitter. The charm mass is set to $m_c = 1.38$~GeV, 
as estimated from HERA charm production cross section~\cite{Abramowicz:1900rp}, 
and the bottom mass to $m_b = 4.75$~GeV. 
The strong-interaction coupling constant at the $Z$ boson mass,
$\alpha_s(M_Z)$, is set to 0.118, and two-loop order is used for the running of $\alpha_s$.

The PDFs for the gluon, $u$-valence, $d$-valence, $\bar{u}$, $\bar{d}$
quark densities are parametrised at the input scale of
$Q^2_0=1.7$~GeV$^2$.
The contribution of the $s$-quark density is taken to be proportional to the
$\bar{d}$-quark density by setting $x\bar{s}(x) = r_s x\bar{d}(x)$,
with $r_s=1.0$. The strange and
anti-strange quark densities are taken to be equal: $x\bar{s}(x) = x
s(x)$.

The impact of a new data set on a given PDF set can be quantitatively
estimated with a profiling procedure~\cite{Paukkunen:2014zia}. The
profiling is performed using a $\chi^2$ function which includes both
the experimental uncertainties and the theoretical uncertainties
arising from PDF variations:

\begin{eqnarray}
\chi^2(\boldsymbol{\beta_{\rm exp}},\boldsymbol{\beta_{\rm th}}) =
 \sum_{i=1}^{N_{\rm data}} \frac{\textstyle \left( \sigma^{\rm exp}_i + \sum_j \Gamma^{\rm exp}_{ij} \beta_{j,\rm exp} - \sigma^{\rm th}_i - \sum_k \Gamma^{\rm th}_{ik}\beta_{k,\rm th} \right)^2}{\Delta_i^2}
 + \sum_j \beta_{j,\rm exp}^2 + \sum_k \beta_{k,\rm th}^2\,.   \label{eq:chi2prof}
\end{eqnarray}
The correlated experimental and theoretical uncertainties are included
using the nuisance parameter vectors $\boldsymbol{\beta_{\rm exp}}$
and $\boldsymbol{\beta_{\rm th}}$, respectively. Their influence on
the data and theory predictions is described by the $\Gamma^{\rm
exp}_{ij}$ and $\Gamma^{\rm th}_{ik}$ matrices. The index $i$ runs
over all $N_{\rm data}$ data points, whereas the index $j$ ($k$)
corresponds to the experimental (theoretical) uncertainty nuisance
parameters. The measurements and the uncorrelated experimental
uncertainties are given by $\sigma^{\rm exp}_i$ and $\Delta_i$\,, respectively, and
the theory predictions are $\sigma_i^{\rm th}$.

\section{Results}
A QCD fit analysis performed on the Tevatron $W$- and $Z$-boson data, together with the HERA I
data, is used to assess the impact of the Tevatron data on
PDFs. The profiling is used to assess the impact of the Tevatron data on various PDF sets.

The optimal functional form for the PDF fit, which corresponds to 15 free
parameters, is found through a \emph{parametrisation scan}, and is used for a fit to
the HERA I data only, and for a fit to the HERA I and Tevatron $W$- and
$Z$-boson data. Table~\ref{tab:chi2fit} shows the
$\chi^2_{\textrm{min}}$ per degrees of freedom (dof) of the fit to the HERA I and Tevatron $W$- and
$Z$-boson data. The contribution to the
total $\chi^2_{\textrm{min}}$ of each data set, referred
to as \emph{partial} $\chi^2$, is also shown.
The partial $\chi^2$ per number of points of each of the Tevatron and HERA I
data set is close to unity.
\begin{table}
  \begin{center}
    \caption{\label{tab:chi2fit} Results of a 15-parameters fit to the
    to the HERA I and Tevatron $W$- and $Z$-boson
    data. The contribution to the total $\chi^2_{\textrm{min}}$ of
    each data set and the corresponding number of points are shown.}
    \begin{tabular}{lc}
      \toprule
      Data set        & $\chi^2$ / number of points \\
      \midrule
      HERA I & 515 / 550 \\
      D0 $d\sigma(Z)/dy$   & 23 / 28  \\
      CDF $d\sigma(Z)/dy$  & 33 / 28  \\
      D0 muon charge asymmetry in $W \to \mu \nu$ & 12 / 10  \\
      CDF $W$ charge asymmetry in $W \to e \nu$  & 15 / 13  \\
      D0 $W$ charge asymmetry in  $W \to e \nu$  & 16 / 14  \\
      \midrule
      Total $\chi^2_{\textrm{min}}$ / dof  & 615 / 628  \\
      \bottomrule
    \end{tabular}
  \end{center}
\end{table}

\begin{figure*}
    \begin{center}
    \subfigure[]{\includegraphics[width=0.4\textwidth]{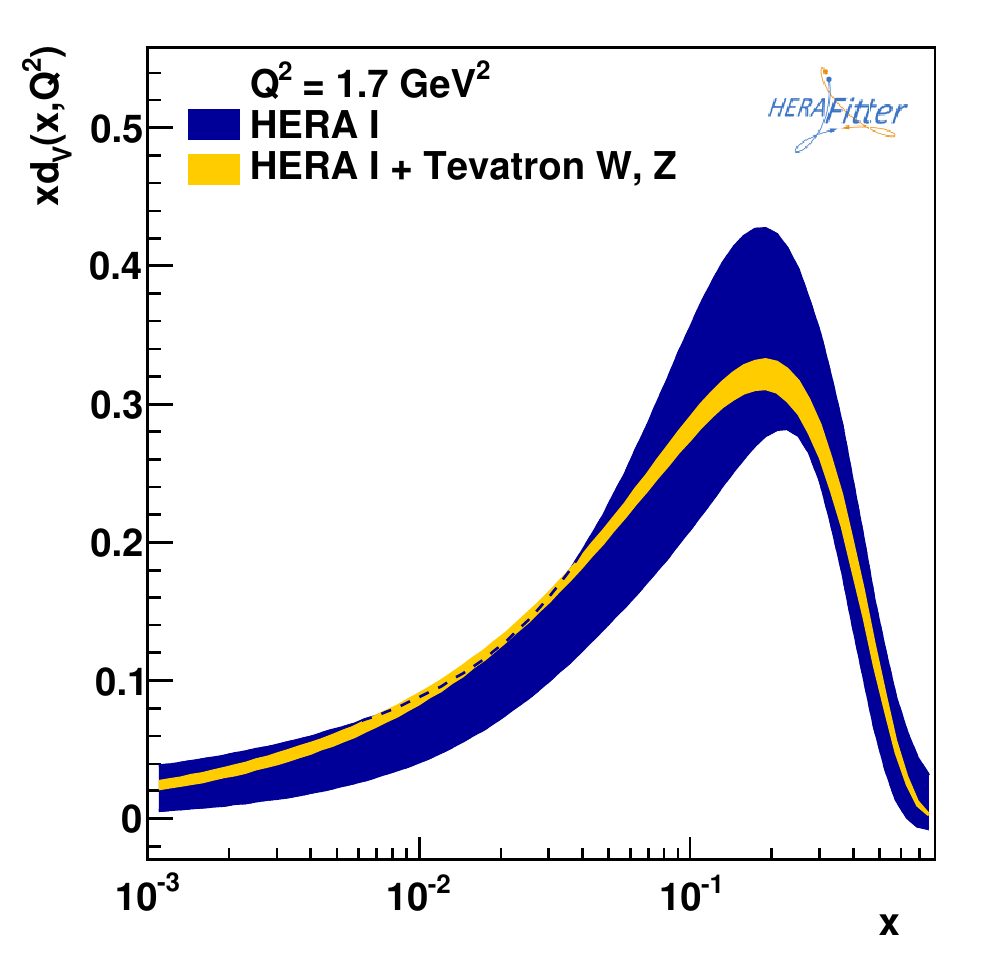}}
    \subfigure[]{\includegraphics[width=0.4\textwidth]{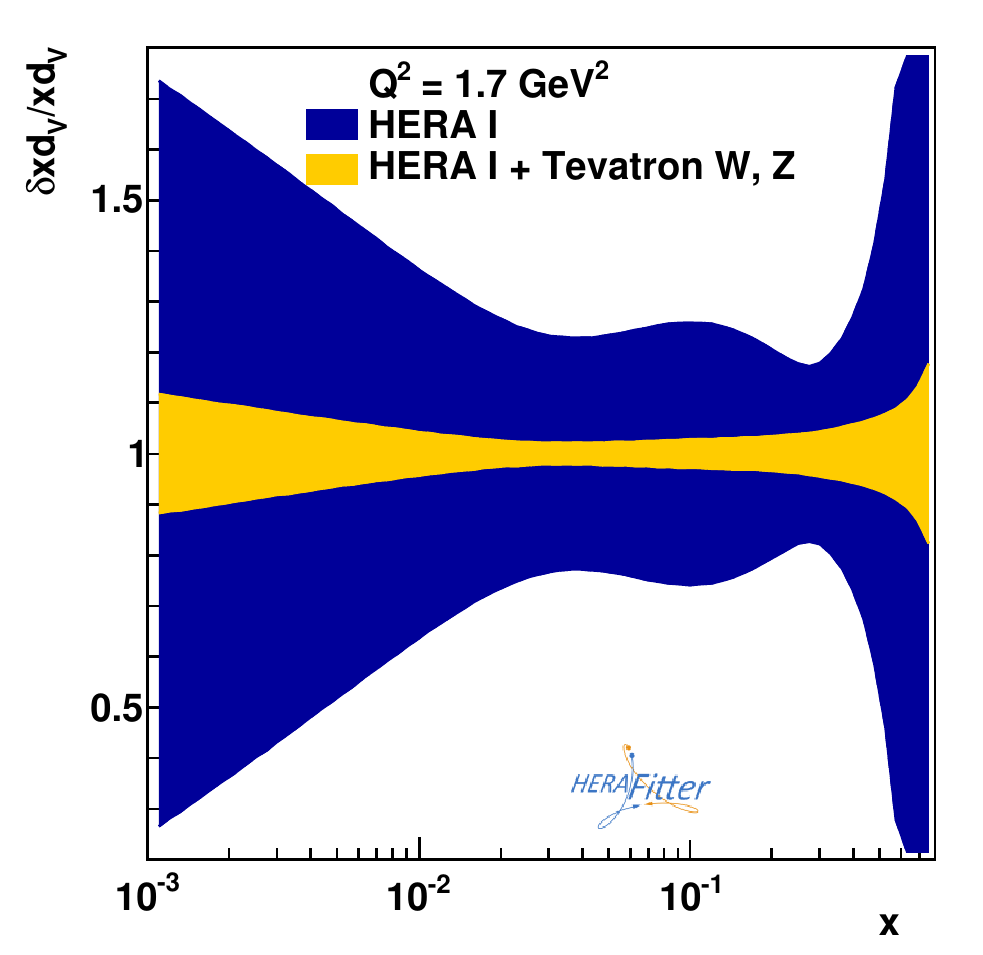}}
  \end{center}
  \caption{\label{fig:pdffit} (a) PDFs at the starting scale $Q^2
    =1.7$~GeV$^2$ as a function of Bjorken-$x$ for $d_v$ determined with a fit
    to the HERA I data (blue), and with a fit to the HERA I and
    Tevatron $W$- and $Z$-boson data (yellow). (b) Relative PDF uncertainties.
  }
\end{figure*}

The central value and the uncertainties of the PDFs are evaluated with
MC replicas~\cite{Forte:2002fg}. Figure~\ref{fig:pdffit} shows the comparison
of the PDFs extracted with the MC replica method by fitting the HERA I
data, and by fitting the HERA I and Tevatron $W$- and $Z$-boson
data. A significant reduction of the PDF
uncertainties is observed in the fit which includes the Tevatron $W$-
and $Z$-boson measurements, in particular for the $d$-valence quark.

The impact of the Tevatron $W$- and $Z$-boson measurements on the
CT10nlo~\cite{Gao:2013xoa} and MMHT2014~\cite{Harland-Lang:2014zoa} is
assessed by profiling.
The uncertainties of the CT10nlo PDFs are scaled to 68\% confidence limit.
The compatibility of the Tevatron data with the
CT10nlo, MMHT2014 and NNPDF3.0~\cite{Ball:2014uwa} sets is tested by evaluating the
$\chi^2$ function of Eq.~(\ref{eq:chi2prof}).
The partial $\chi^2$ per number of points
of each of the Tevatron data set, and the total $\chi^2$ / dof, are
close to unity for all the PDFs.
 
\begin{figure*}
  \begin{center}
    \subfigure[]{\includegraphics[width=0.4\textwidth]{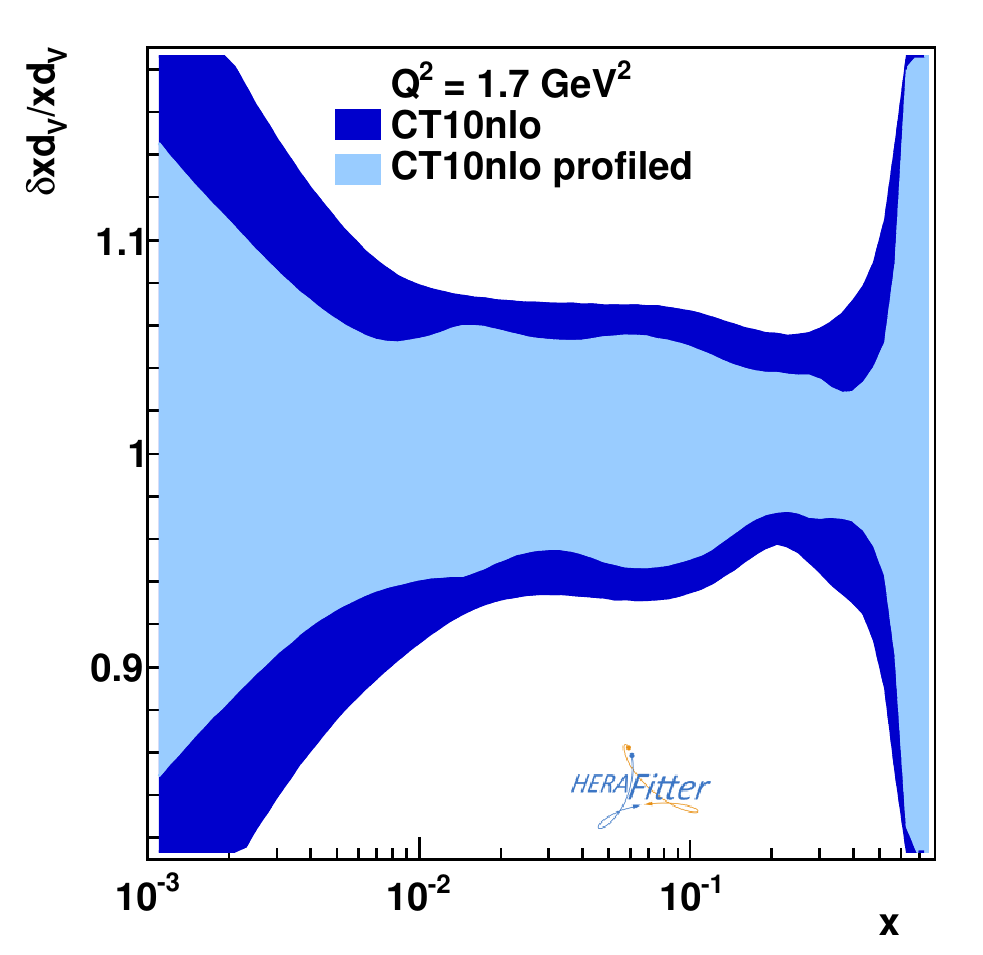}}
    \subfigure[]{\includegraphics[width=0.4\textwidth]{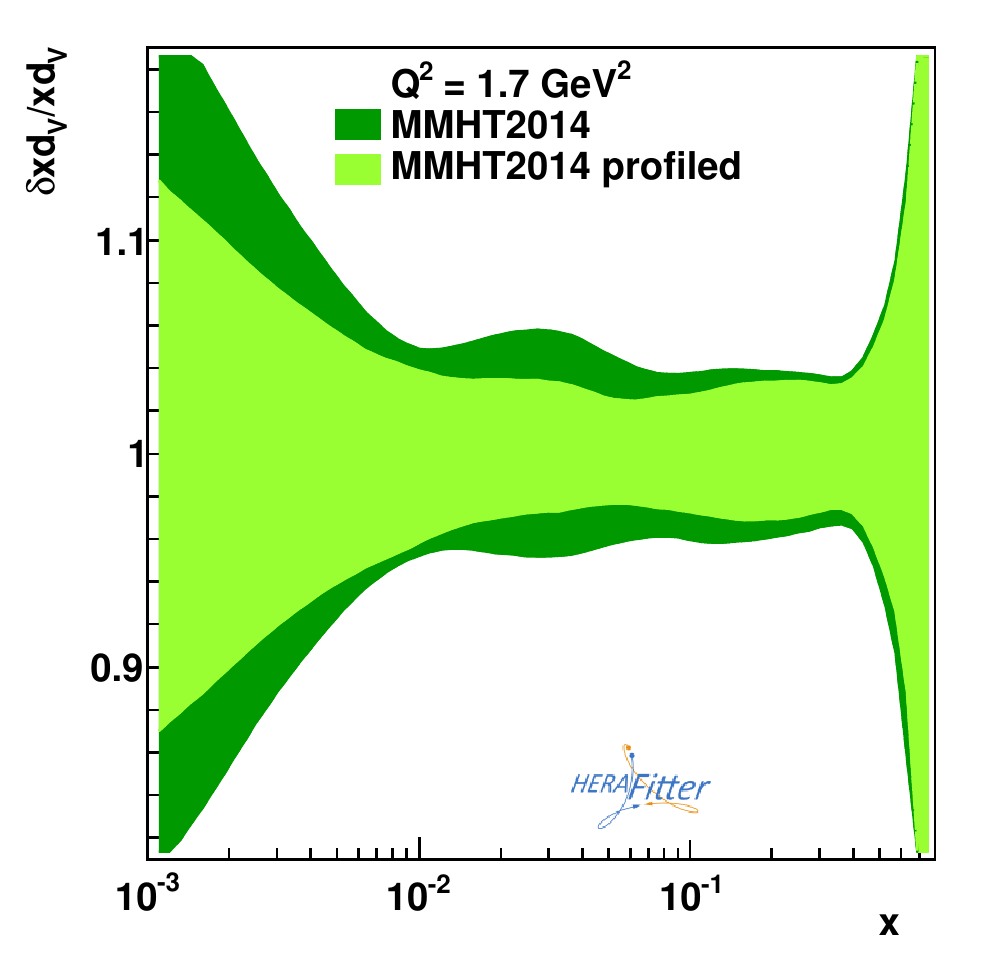}}
    \caption{Relative uncertainties of the $d$-valence PDF at the scale $Q^2 = 1.7$~GeV$^2$ as a function of Bjorken-$x$ before
    and after profiling for the (a) CT10nlo and (b) MMHT2014 PDFs.
\label{fig:pdfprofiled}}
  \end{center}
\end{figure*}

The CT10nlo and MMHT2014 PDFs are profiled to the Tevatron $W$- and
$Z$-boson data. The results of the profiling on the relative uncertainty of the $d$-valence
PDF are shown in Fig.~\ref{fig:pdfprofiled}. Significant reduction of the uncertainty is observed for both
sets.

Tables of the Tevatron measurements, with updated correlation model,
and corresponding APPLGRID theoretical predictions~\cite{Carli:2010rw} are publicly
available at \emph{herafitter.org}.

\section*{References}

\bibliographystyle{utphys_mod}
\bibliography{camarda_herafitter}

\end{document}